\documentclass[12pt]{article}
\usepackage{latexsym,epsfig,graphicx,a4wide,amssymb}

\def\b{\beta}

\def\d{\delta}

{\rm }

\def\s{\sigma}

\def\be{\begin{equation}}
 \def\ee{\end{equation}}
 \def\bea{\begin{eqnarray}}
 \def\eea{\end{eqnarray}}
 
 \def\b{\beta}
 
 \def\d{\delta}
 
 \def\s{\sigma}

\def\L{\Lambda}

\newcommand{\fr}{\frac}
\newcommand{\pr}{\prime}

\def\pa{\partial}

\def\2{\frac{1}{2}}
\def\4{\frac{1}{4}}

\def\PLA#1{{Phys.\ Lett.\ A \bf #1}}
\def\PR#1{{Phys.\ Rev.\ D \bf #1}}
\def\PRL#1{{Phys.\ Rev.\ Lett.\ \bf #1}}
\def\cm#1{{Commun.\ Math.\ Phys.\ \bf #1}}

\def\cqg#1{{Class.\ Quant.\ Grav.\ \bf #1}}
\def\gen{\mathrm{g}}

\catcode`\@=11

\@addtoreset{equation}{section}

\def\@normalsize{\@setsize\normalsize{15pt}\xiipt\@xiipt
\abovedisplayskip 14pt plus3pt minus3pt%
\belowdisplayskip \abovedisplayskip
\abovedisplayshortskip  \z@ plus3pt%
\belowdisplayshortskip  7pt plus3.5pt minus0pt}
\def\small{\@setsize\small{13.6pt}\xipt\@xipt
\abovedisplayskip 13pt plus3pt minus3pt%
\belowdisplayskip \abovedisplayskip
\abovedisplayshortskip  \z@ plus3pt%
\belowdisplayshortskip  7pt plus3.5pt minus0pt
\def\@listi{\parsep 4.5pt plus 2pt minus 1pt
            \itemsep \parsep
            \topsep 9pt plus 3pt minus 3pt}}

\def\underline#1{\relax\ifmmode\@@underline#1\else
        $\@@underline{\hbox{#1}}$\relax\fi}
\@twosidetrue \relax

\catcode`@=12

\catcode`\@=11

\def\section{\@startsection{section}{1}{\z@}{3.5ex plus 1ex minus
   .2ex}{2.3ex plus .2ex}{\large\bf}}

\def\ps@headings{\def\@oddfoot{}\def\@evenfoot{}
\def\@oddhead{\hbox{}\hfill
        \makebox[.5\textwidth]{\raggedright\ignorespaces --\thepage{}--
        \hfill }}
\def\@evenhead{\@oddhead}
\def\subsectionmark##1{\markboth{##1}{}}
}

\ps@headings

\catcode`\@=12

\begin{document}

\begin{titlepage}

\rightline{UTHET-09-1001}

\begin{centering}
\vspace{1cm}
{\Large {\bf A New Class of Exact Hairy  Black Hole Solutions
}}\\

\vspace{1.5cm}

 {\bf Theodoros Kolyvaris $^{\dagger}$}, {\bf George Koutsoumbas $^{\sharp}$},\\ {\bf Eleftherios Papantonopoulos $^{*}$} \\
 \vspace{.2in}
 Department of Physics, National Technical University of
Athens, \\
Zografou Campus GR 157 73, Athens, Greece \\
\vspace{.2in}
 {\bf George Siopsis $^{\flat}$}
\vspace{.2in}

 Department of Physics and Astronomy, The
University of Tennessee,\\ Knoxville, TN 37996 - 1200, USA
 \\
\vspace{3mm}

\end{centering}
\vspace{1.5cm}

\begin{abstract}

We present a new class of black hole solutions  with a minimally
coupled scalar field in the  presence of a negative cosmological
constant. We consider an one-parameter family of self-interaction
potentials parametrized by a dimensionless parameter $g$. When
$g=0$, we recover the conformally invariant solution of  the
Martinez-Troncoso-Zanelli (MTZ) black hole. A non-vanishing $g$
signals the departure from conformal invariance.
Thermodynamically, there is a critical temperature at vanishing
black hole mass, where a higher-order phase transition occurs, as
in the case of the MTZ black hole. Additionally, we obtain a
branch of hairy solutions which undergo a first-order phase
transition at a second critical temperature which depends on $g$
and it is higher than the MTZ critical temperature. As $g\to 0$,
this second critical temperature diverges.

\end{abstract}

\vspace{1.5cm}
\begin{flushleft}
$^{\dagger}~~$ e-mail address: teokolyv@central.ntua.gr \\
$^{\sharp}~~$ e-mail address: kutsubas@central.ntua.gr \\
$^{*} ~~$ e-mail address: lpapa@central.ntua.gr \\
$ ^{\flat}~~$ e-mail address: siopsis@tennessee.edu

\end{flushleft}
\end{titlepage}

\section{Introduction}

Four-dimensional black hole solutions of Einstein gravity coupled
to a scalar field have been an avenue of intense research for many years.
Questions pertaining to their existence, uniqueness and stability were
seeking answers over these years. The Kerr-Newman solutions of
four-dimensional asymptotically flat black holes coupled to an
electromagnetic field or in vacuum, imposed very stringent
conditions on their existence in the form of ``no-hair"
theorems. In the case of a minimally coupled scalar field in
asymptotically flat spacetime the no-hair theorems were proven
imposing conditions on the form of the self-interaction
potential~\cite{nohairtheo}. These theorems were also generalized
to non-minimally coupled scalar fields~\cite{Mayo}.

For asymptotically flat spacetime, a four-dimensional black hole
coupled to a scalar field with a zero self-interaction potential
is known~\cite{BBMB}. However, the scalar field diverges on the
event horizon and, furthermore, the solution is unstable
\cite{bronnikov}, so there is no violation of the ``no-hair"
theorems.   In the case of a positive cosmological constant with a
minimally coupled scalar field with a self-interaction potential,
black hole solutions were found in~\cite{Zloshchastiev:2004ny} and
also a numerical solution was presented in~\cite{Torii:1998ir},
but it was unstable. If the scalar field is non-minimally coupled,
a solution exists with a quartic self-interaction
potential~\cite{martinez}, but it was shown to be
unstable~\cite{phil,dotti}.

In the case of a negative cosmological constant, stable solutions
were found numerically for spherical
geometries~\cite{Torii:2001pg, Winstanley:2002jt} and an exact
solution in asymptotically AdS space with hyperbolic geometry was
presented in~\cite{Martinez:2004nb} and generalized later to
include charge~\cite{Martinez:2005di}. This solution is
perturbatively stable for negative mass and may develop
instabilities for positive mass~\cite{papa1}. The thermodynamics
of this solution were studied in~\cite{Martinez:2004nb} where it
was shown that there is a second order phase transition of the
hairy black hole to a pure topological black hole without hair.
The analytical and numerical calculation of the quasi-normal modes
of scalar, electromagnetic and tensor perturbations of these black
holes confirmed this behaviour~\cite{papa2}. Recently, a new exact
solution of a charged C-metric conformally coupled to a scalar
field was presented in~\cite{kolyvaris, anabalon}. A
Schwarzschild-AdS black hole in five-dimensions coupled to a
scalar field was discussed in~\cite{Farakos:2009fx}, while
Dilatonic black hole solutions with a Gauss-Bonnet term in various
dimensions were discussed in~\cite{Ohta:2009pe}.

From a known black hole solution coupled to a scalar field other
solutions  can be generated  via conformal
mappings~\cite{conformal}. In all black hole solutions in the
Einstein frame the scalar field is coupled minimally to gravity.
Applying a conformal transformation to these solutions, other
solutions can be obtained in the Jordan frame which are not
physically equivalent to the untransformed ones~\cite{varofone}.
The scalar field in the Jordan frame is coupled to gravity
non-minimally and this coupling is characterized by a
dimensionless parameter $\xi$. There are strong theoretical,
astrophysical and cosmological arguments (for a review
see~\cite{varofone}) which fix the value of  this conformal
coupling to $\xi=1/6$. If the scalar potential is zero or quartic
in the scalar field, the theory is conformally invariant; otherwise a
non-trivial scalar potential introduces a scale in the theory and
the conformal invariance is broken.

In this work we present a new class of black hole solutions of
four-dimensional Einstein gravity coupled to a scalar field and to
vacuum energy. We analyse the structure and study the properties
of these solutions in the Einstein frame. In this frame,  the
scalar self-interaction potential is characterised by a
dimensionless parameter $g$. If this parameter vanishes, then the
known solutions of black holes minimally coupled to a scalar field
in (A)dS space are obtained~\cite{martinez,Martinez:2004nb}.
Transforming these solutions to the Jordan frame, the  parameter
$g$ can be interpreted as giving the measure of departure from
conformal invariance. This breakdown of conformal invariance
allows the back-scattering of waves of the scalar field off of the
background curvature of spacetime, and the creation of ``tails''
of radiation. This effect may have sizeable observation signatures
in cosmology~\cite{waves}.

Following~\cite{Hertog:2004bb}, we perform a perturbative
stability analysis of the solutions. We find that the hairy black
hole is stable near the conformal point if the mass is
negative and may develop instabilities in the case of positive
mass.
We also study the thermodynamics of our solutions. Calculating the
free energy we find that there is a critical temperature above
which the hairy black hole loses its hair to a black hole in
vacuum. This critical temperature occurs at a point where the
black hole mass flips sign, as in the case of the MTZ black hole
\cite{Martinez:2004nb}.  Interestingly, another phase transition
occurs at a higher critical temperature which is of first order
and involves a different branch of our solution. This new critical
temperature diverges as the coupling constant in the potential
$g\to 0$.
These exact hairy black hole
solutions may have interesting applications to holographic
superconductors~\cite{Hartnoll:2008vx,Hartnoll:2008kx}, where new
types of holographic superconductors can be
constructed~\cite{papa1,zeng}.

Our discussion is organized as follows. In section \ref{sec2} we
introduce the self-interaction potential and we present the hairy
black hole solution. In section \ref{sec4} we discuss the
thermodynamics of our solution. In section \ref{secst} we perform
a stability analysis. Finally, in section \ref{sec6} we summarize
our results.

\section{Black Hole with Scalar Hair}
\label{sec2}

To obtain a black hole with scalar hair, we start with the
four-dimensional action consisting of the Einstein-Hilbert action
with a negative cosmological constant
$\Lambda$,
along with a scalar,
\begin{equation}
I = \int d^4x\sqrt{-g}\left[ \fr{R-2\Lambda}{16 \pi G} -\fr{1}{2}
g^{\mu\nu}\pa_\mu\phi\pa_\nu\phi-V(\phi)\right],\label{action11}
\end{equation} where $G$ is Newton's constant and $R$ is the Ricci
scalar. The corresponding field equations are
\begin{eqnarray}
G_{\mu\nu} +\Lambda g_{\mu\nu}&=&8\pi G T_{\mu\nu}^{\mathrm{matter}}~, \nonumber\\
\Box \phi&=&\frac{d V}{d \phi},
\label{eqfe}\end{eqnarray}
where the energy-momentum tensor is given by
\begin{equation}
\label{Tuvfield}T_{\mu\nu}^{\mathrm{matter}} =\pa_\mu \phi
\pa_\nu\phi-\fr{1}{2} g_{\mu\nu} g^{\alpha\beta}\pa_\alpha \phi
\pa_\beta \phi - g_{\mu\nu} V(\phi)~.
\end{equation}
The potential is chosen as
\begin{eqnarray} V(\phi) &=&
\fr{\Lambda}{4\pi G}\sinh^2\sqrt{\fr{4 \pi G}{3}}\phi\nonumber\\
&+& \frac{g\Lambda}{24\pi G} \left[2 \sqrt{3 \pi G} \phi \cosh
\left(\sqrt{\fr{16 \pi G}{3}}\phi\right) - \frac{9}{8} \sinh
\left(\sqrt{\fr{16 \pi G}{3}}\phi\right)- \frac{1}{8} \sinh
\left(4 \sqrt{3 \pi
G}\phi\right)\right]\nonumber\\\label{potentialnew}
\end{eqnarray}
and it is given in terms of a coupling constant $g$. Setting $g=0$
we recover the action that yields the MTZ black hole
\cite{Martinez:2004nb}.
This particular form of the potential is chosen so that the field equations can be solved analytically.
The qualitative nature of our results does not depend on the detailed form of the potential. A similar potential was considered in a different context in \cite{Zloshchastiev:2004ny} (see also \cite{zeng} for the derivation of a potential that yields analytic solutions in the case of a spherical horizon).
If one goes over to the Jordan frame, in which the scalar field obeys the Klein-Gordon equation
\be \label{KleinGordon} \Box \phi -\xi R
\phi -\frac{dV}{d\phi}=0 \; , \ee with $\xi=1/6$, the
scalar potential has the form
\bea V (\phi) = -\frac{2 \pi G \L}{9}~\phi^{4} &-& \frac{g\L}{16 \pi G}
\left[ \sqrt{\frac{16\pi G}{3}}~\phi\left( 1-\frac{4 \pi G}{3}~\phi^{2}
+\frac{\frac{16\pi G}{9}~\phi^{2}}{1-\frac{4 \pi G}{3}~\phi^{2}}\right) \right.\nonumber \\
&-& \left.\left( 1-\frac{4\pi G}{3}~\phi^{2}\right)\left( 1+\frac{4\pi G}{3}~\phi^{2}\right) \ln\frac{1+\sqrt{\frac{4\pi G}{3}}~\phi}{1-\sqrt{\frac{4 \pi G}{3}}~\phi}
\right]~. \label{potentialnm}
\eea
Evidently, the scalar field is
conformally coupled but the conformal invariance is broken by a
non-zero value of $g$.

The mass of the scalar field is given by
\be m^2 = V''(0) = - \frac{2}{l^2} \ee
where we defined $\Lambda = -3/l^2$.
Notice that it is independent of $g$ and coincides with the scalar mass that yields the MTZ black hole
\cite{Martinez:2004nb}.
Asymptotically ($r\to\infty$), the scalar field
behaves as $\phi \sim r^{-\Delta_\pm}$ where $\Delta_\pm = \frac{3}{2} \pm \sqrt{ \frac{9}{4} + m^2l^2}$. In our case $\Delta_+ = 2$ and $\Delta_- = 1$.
Both boundary conditions are acceptable as both give normalizable modes.
We shall adopt the mixed boundary conditions (as $r\to\infty$) \be \phi
(r) = \frac{\alpha}{r} + \frac{c\alpha^2}{r^2} + \dots \ , \ \ \ \
c = -\sqrt{\frac{4\pi G}{3}} < 0~. \label{eqBCphi}\ee This choice of the parameter $c$ coincides with the MTZ solution
\cite{Martinez:2004nb}.

Solutions to
the Einstein equations with the boundary conditions (\ref{eqBCphi}) have been
found in the case of spherical horizons and shown to be unstable
\cite{Hertog:2004bb}. In that case, for $\alpha >0$, it was shown
that $c<0$ always and the hairy black hole had positive mass. On
the other hand, MTZ black holes, which have hyperbolic horizons
and obey the boundary conditions (\ref{eqBCphi}) with $c<0$, can
be stable if they have negative mass \cite{papa1}. This is
impossible with spherical horizons, because they always enclose
black holes of positive mass. The numerical value of $c$ is not
important (except for the fact that $c\ne 0$) and is chosen as in
(\ref{eqBCphi}) for convenience.

The field equations admit solutions which are black holes with topology
$\mathbb{R}^{2}\times\Sigma$, where $\Sigma$ is a two-dimensional
manifold of constant negative curvature. Black holes with constant
negative curvature are known as topological black holes (TBHs - see, e.g., \cite{bibMann, bibBir}).
The simplest solution for $\Lambda=-3/l^2$ reads
\begin{equation}\label{linel}
ds^{2}=-f_{TBH}(\rho)dt^{2}+\frac{1}{f_{TBH}(\rho)}d\rho^{2}
+\rho^{2}d\sigma ^{2}\quad ,\quad
f_{TBH}(\rho)=\frac{\rho^{2}}{l^2} -1-\frac{\rho_0 }{\rho}~,
\end{equation}
where
$\rho_0$ is a constant which is proportional to the mass and is
bounded from below ($\rho_0\geq-\frac{2}{3\sqrt{3}} l$),
$d\sigma^{2}$ is the line element of the two-dimensional manifold
$\Sigma$ which is locally isomorphic to the hyperbolic manifold
$H^{2}$ and of the form
\begin{equation}\label{eqSigma}
\Sigma=H^{2}/\Gamma \quad \textrm{,\quad  $\Gamma\subset
O(2,1)$}~,
\end{equation}
with $\Gamma$  a freely acting discrete subgroup (i.e., without
fixed points) of isometries. The line element $d\sigma^{2}$ of
$\Sigma$ can be written as
\begin{equation}
d\sigma^{2}=d\theta^{2}+\sinh^{2}\theta d\varphi{^2}~,
\end{equation}
with $\theta\ge0$ and $0\le\phi<2\pi$ being the coordinates of the
hyperbolic space $H^{2}$ or pseudosphere, which is a non-compact
two-dimensional space of constant negative curvature. This space
becomes a compact space of constant negative curvature with genus
$\gen\ge2$ by identifying, according to the connection rules of
the discrete subgroup $\Gamma$, the opposite edges of a
$4\gen$-sided polygon whose sides are geodesics and is centered at
the origin $\theta=\varphi=0$ of the pseudosphere. An octagon is
the simplest such polygon, yielding a compact surface of genus
$\gen=2$ under these identifications. Thus, the two-dimensional
manifold $\Sigma$ is a compact Riemann 2-surface of genus
$\mathrm{g}\geq2$.  The configuration (\ref{linel}) is an
asymptotically locally AdS spacetime. The horizon structure of
(\ref{linel}) is determined by the roots of the metric function
$f_{TBH}(\rho)$, that is
\begin{equation}
f_{TBH}(\rho)=\frac{\rho^{2}}{l^2}-1-\frac{\rho_0}{\rho}=0~.
\label{ftbh}
\end{equation}
For $-\frac{2}{3\sqrt{3}} l <\rho_0<0$, this equation has two
distinct non-degenerate solutions, corresponding to an inner and
to an outer horizon $\rho_{-}$ and $\rho_{+}$ respectively. For
$\rho_0\geq0$, $f_{TBH}(\rho)$ has just one non-degenerate root
and so the black hole (\ref{linel}) has one horizon $\rho_{+}$.
The horizons for both cases of $\rho_0$ have the non-trivial
topology of the manifold $\Sigma$. We note that for
$\rho_0=-\frac{2}{3\sqrt{3}} l$, $f_{TBH}(\rho)$ has a degenerate
root, but this horizon does not have an interpretation as a black
hole horizon.

The boundary has the metric \be\label{eqnew} ds_\partial^2 = -dt^2
+ l^2 d\sigma^2~, \ee so spatially it is a hyperbolic manifold of
radius $l$ (and of curvature $-1/l$).

The action (\ref{action11}) with a potential as in
(\ref{potentialnew}) has a static black hole solution with
topology $\mathbb{R}^{2}\times \Sigma $ and with scalar hair, and
it is given by
\begin{equation}\label{MTZconf}
ds^2=\frac{r (r+2r_0)}{(r+r_0)^2} \left[-F(r)
dt^2+\frac{dr^2}{F(r)}+r^2 d \sigma^2\right]~, \ee where
\be\label{Psiconf} F(r) = \frac{r^2}{l^2} - g\frac{r_0}{l^2} r - 1
+ g \frac{r_0^2}{l^2} - \left( 1 - 2g \frac{r_0^2}{l^2} \right)
\frac{r_0}{r} \left( 2 + \frac{r_0}{r} \right) +g \frac{r^2}{2l^2}
\ln\left( 1 + \frac{2r_0}{r}\right)\ , \ee and the scalar field is
 \be \phi(r) = \sqrt{\fr{3}{4 \pi G}}~ {\rm arctanh}
\left(\fr{r_0}{r+r_0}\right)~, \label{scaField} \ee
obeying the boundary conditions (\ref{eqBCphi}) by design.


\section{Thermodynamics}
\label{sec4}

To study the thermodynamics of our black hole solutions we
consider the Euclidean continuation ($t \rightarrow i\tau$) of the
action in Hamiltonian form
\begin{equation}
I=\int \left[ \pi ^{ij}\dot{g}_{ij}+p\dot{\phi}-N\mathcal{H}-N^{i}\mathcal{H}%
_{i}\right] d^{\,3}xdt+B,  \label{Hamiltaction}
\end{equation}
where $\pi ^{ij}$ and $p$ are the conjugate momenta of the metric
and the field respectively; $B$ is a surface term. The solution
reads:
\begin{equation}
ds^{2}=N^{2}(r)f^{\,2}(r)d\tau^{2}+f^{\,-2}(r)dr^{2}+R^{2}(r)d\s
^{2} \label{metricEucl}
\end{equation}
where
\be
    N(r)    =  \fr{r(r+2r_0)}{(r+r_0)^2}  \ , \ \
    f^2(r)  =  \fr{(r+r_0)^2}{r(r+2r_0)}\;F(r) \ , \ \
    R^2(r)  =  \fr{r^3(r+2r_0)}{(r+r_0)^2}~,
\ee
with a periodic $\tau$ whose period is the inverse temperature, $\beta = 1/T$.

The Hamiltonian action becomes
\begin{equation}
I=-\b \,\s \int_{r_{+}}^{\infty }N(r)\mathcal{H}(r)dr+B,
\label{redHamiltaction}
\end{equation}
where $\sigma $ is the area of $\Sigma $ and
\be
\mathcal{H} =NR^{2}\left[ \frac{1}{8\pi G}\left(
\frac{(f^{\,2})^{\pr }R^{\,\pr}}{R}+\frac{2f^{\,2}R^{\,\pr \pr
}}{R}+\frac{1}{R^{\,2}}(1
+f^{\,2})
+\L \right) +\frac{1}{2}f^{\,2}(\phi ^{\pr})^{2}+V(\phi )\right]
.
\ee
The Euclidean solution is static and satisfies the equation $%
\mathcal{H}=0$. Thus, the value of the action in the classical
limit is just the surface term $B$, which should maximize the
action within the class of fields considered.

We now compute the action when the field equations hold. The
condition that the geometries which are permitted should not have
conical singularities at the horizon imposes
\be
 T =  \fr{F^{\;\pr}(r_+)}{4\pi} \label{eqtemperature}\,.
\ee
Using the grand canonical ensemble (fixing the temperature),
the variation of the surface term reads
\[
\d B\equiv \d B_{\phi }+\d B_{G}\;,
\]
where
\begin{equation}
\d B_{G}=\frac{\b \s }{8\pi G} \Big{[} N \Big{(} RR^{\,\pr }\d
f^{\,2}-(f^{\,2})^{\pr }R\d R\Big{)} +2f^{\,2}R\Big{(} N\d
R^{\,\pr }-N^{\pr }\d R\Big{)} \Big{]} _{r_{+}}^{\infty }\;,
\label{delG}
\end{equation}
and the contribution from the scalar field equals
\begin{equation}
\d B_{\phi }=\b \s NR^{\,2}f^{\,2}\phi ^{\pr }\d \phi
\big|_{r_{+}}^{\infty }\;.  \label{delphi}
\end{equation}
For the metric, the variation of fields at infinity yields
\begin{eqnarray}
\left. \d f^{\,2}\right| _{\infty } &=& \left(\fr{2}{l^2}r_0-\fr{2(3+(9-8g) r_0^2/l^2)}{3r}-\fr{4r_0(1-4 r_0^2/l^2)}{r^2}
+\mathcal{O}(r^{-3})\right)\d r_0~,\nonumber\\
 \d \phi \big| _{\infty } &=& \sqrt{\fr{3}{4\pi G}}\left(\fr{1}{r}-\fr{2r_0}{r^2}+\mathcal{O}(r^{-3})\right)\d r_0~,\nonumber\\
 \d R\big| _{\infty } &=& \left(-\fr{r_0}{r}+\fr{3r_0^2}{r^2}+\mathcal{O}(r^{-3})\right)\d r_0~,
\end{eqnarray}
so
\begin{eqnarray}
 \d B_{G}\big| _{\infty } &=& \frac{\b\s}{8\pi G}\left( \frac{6 r_0( r-4(1-2g/9)r_0)}{l^2} -2+\mathcal{O}(r^{-1})\right) \d r_0\;, \nonumber\\
 \delta B_{\phi }\big| _{\infty } &=& \frac{\b\s}{8\pi G}\Big( -\frac{6 r_0(r-4r_0)}{l^2}+\mathcal{O}(r^{-1})\Big) \d r_0\;.
\label{B1}
\end{eqnarray}
The surface term at infinity is
\begin{equation}
     B\big| _{\infty }=-\frac{\b\s(3-8gr_0^2/l^2)}{12\pi G} r_0\;.  \label{Binf}
\end{equation}
The variation of the surface term at the horizon may be found using
the relations
\begin{eqnarray*}
\left. \d R\right| _{r_{+}} &=&\d R(r_{+})-\left. R^{\,\pr
}\right|
_{r_{+}}\d r_{+}\;, \\
\left. \d f^{\,2}\right| _{r_{+}} &=&-\left. (f^{\,2})^{\pr
}\right| _{r_{+}}\d r_{+}\;.
\end{eqnarray*}
We observe that $\left.
\delta B_{\phi }\right| _{r_{+}}$ vanishes, since $f^2(r_+)=0$,
and
\begin{eqnarray*}
\left. \d B\right| _{r_{+}} &=&-\frac{\b \s}{16\pi G}%
N(r_{+})\left. (f^{\,2})^{\pr }\right| _{r_{+}}\d R^{\,2}(r_{+}) \\
&=&-\frac{\s}{4 G}\d R^{\,2}(r_{+})\;.
\end{eqnarray*}
Thus the surface term at the horizon is
\begin{equation}
\left. B\right| _{r_{+}}=-\frac{\s }{4 G}R^{\,2}(r_{+})\;.
\label{Bhor}
\end{equation}
Therefore,
provided the field equations hold, the Euclidean action reads
\begin{equation}
I=-\frac{\b\s(3-8gr_0^2/l^2)}{12\pi G} r_0 +\frac{\s}{4 G}R^{\,2}(r_{+})\;.
\label{Ionshell}
\end{equation}
The Euclidean action is related to the free energy through $I=-\b
F$. We deduce
\begin{equation}
I=S-\b M\;,  \label{IMS}
\end{equation}
where $M$ and $S$ are the mass and entropy respectively,
\be
    M = \fr{\s(3-8gr_0^2/l^2)}{12\pi G}\; r_0 \ , \ \
    S= \fr{\s}{4 G}\;R^2(r_+)=\fr{A_H}{4 G} \label{massentropy}
\ee
It is easy to show that the law of thermodynamics $dM=TdS$ holds.
For $g = 0$, these expressions reduce to the corresponding quantities for MTZ black holes \cite{Martinez:2004nb}.
Alternatively, the mass of the black hole can be found by the Ashtekar-Das method \cite{bibAshDas}. A straightforward calculation confirms the expression (\ref{massentropy}) for the mass.

In the case of the topological black hole (\ref{ftbh}) the
temperature, entropy and mass  are given by respectively, \be
T=\frac{3}{4 \pi l} \left( \frac{\rho_{+}}{l} - \frac{l}{3\rho_+}
\right)~, \quad S_{TBH}=\frac{\sigma \rho^{2}_{+}}{4G}~,\quad
M_{TBH}=\frac{\sigma\rho_{+}}{8 \pi G} \left( \frac{\rho_+^2}{l^2}
- 1 \right)~,\label{relations2} \ee and also  the law of
thermodynamics $dM=TdS$ is obeyed.

We note that, in the limit $r_0\to 0,$ $F(r) \to
\frac{r^2}{l^2} - 1$ from eq.~(\ref{Psiconf}) and the corresponding temperature (\ref{eqtemperature}) reads
$T=\frac{1}{2 \pi l},$ which equals the temperature of the
topological black hole (\ref{relations2}) in the limit $\rho_0\to 0$ ($\rho_+\to 1$). The common limit
\begin{equation}
ds_{\mathrm{AdS}}^{2}=-\left[ \frac{r^{2}}{l^2} -1\right]
dt^{2}+\left[ \frac{r^{2}}{l^2}-1\right] ^{-1}dr ^{2}+r
^{2}d\sigma ^{2}\; \label{muzero}\end{equation} is a manifold of
negative constant curvature possessing an event horizon at $r=l$.
The TBH and our hairy black hole solution match continuously at
the critical temperature \be\label{eqTcr} T_0 = \frac{1}{2\pi l}~,
\ee which corresponds to $M_{TBH} = M = 0$, with (\ref{muzero}) a
transient configuration. Evidently, at the critical point
(\ref{eqTcr}) a scaling symmetry emerges owing to the fact that
the metric becomes pure AdS.

At the critical temperature (\ref{eqTcr}) a higher-order phase
transition occurs as in the case  of the MTZ black hole (with
$g=0$). Introducing the terms with $g\ne 0$ in the potential do
not alter this result qualitatively.

Next we perform a detailed analysis of thermodynamics and examine
several values of the coupling constant $g$. Henceforth we shall
work with units in which $l=1$.
We begin with a geometrical characteristic of the hairy black
hole, the horizon radius $r_+$ (root of $F(r)$
(eq.~(\ref{Psiconf})). In figure \ref{horizons} we show the $r_0$
dependence of the horizon for representative values of the
coupling constant, $g=3$ and $g=0.0005.$ We observe that, for
$g=3,$ the horizon may correspond to more than one value of the
parameter $r_0.$ For $g=0.0005$ we see that, additionally, there
is a maximum value of the horizon radius.

We note that one may express the radius of the horizon $r_{+}$ in
terms of the dimensionless parameter
 \be \xi = \frac{r_0}{r_+} \ee
as \be r_+ = \frac{1+\xi }{\sqrt{1+g \xi (1+\xi)(-1+2\xi + 2\xi^2)
+ \frac{1}{2} g \ln (1+2\xi)}} \label{rhm3}~. \ee The temperature
reads \be T = \frac{1+\xi (1+\xi) (4-g(1+2\xi + 2\xi^2)) +
\frac{1}{2} g (1+2\xi)^2 \ln (1+2\xi)}{2\pi (1+2\xi) \sqrt{1+g \xi
(1+\xi)(-1+2\xi + 2\xi^2) + \frac{1}{2} g \ln (1+2\xi)}}~,
\label{Tm3} \ee or equivalently \be T=\frac{(r_++r_0)(r_+^2+4 r_0
r_+ +4 r_0^2-8 g r_0^3 r_+-8 g r_0^4)}{2 \pi r_+^3}~,\ee a third
order equation in $r_+,$ showing that, for a given temperature,
there are in general three possible values of $\xi$. Thus we
obtain up to three different branches of our hairy black hole
solution.

We start our analysis with a relatively large value of the
coupling constant $g$, namely $g=3$ and calculate the horizon
radius, temperature and Euclidean action for various values of
$r_0$. In figure \ref{mass_action_q3}, left panel, we depict $r_0$
versus $T$ and it is clear that there is a $T$ interval for which
there are really three corresponding values for $r_0.$ Outside
this interval, there is just one solution. The corresponding graph
for the Euclidean actions may be seen in the right panel of the
same figure. The action for the topological black hole with the
common temperature $T$ is represented by a continuous line, while
the actions for the hairy black holes are shown in the form of
points. We note that equation (\ref{ftbh}) yields for the
temperature of the topological black hole $T=\frac{1}{4
\pi}\left(\frac{3 \rho_+}{l^2}+\frac{1}{\rho_+}\right) \Rightarrow
\rho_+=\frac{2 \pi T}{3}+\sqrt{\left(\frac{2 \pi
T}{3}\right)^2-\frac{1}{3}}.$ The largest Euclidean action
(smallest free energy) will dominate.

There are three branches for the hairy black hole, corresponding
to the three different values of $r_0.$ In particular, for some
fixed temperature (e.g., $T=0.16$) the algebraically lowest $r_0$
corresponds to the algebraically lowest Euclidean action;
similarly, the medium and largest $r_0$ parameters correspond to
the medium and largest Euclidean actions. The medium Euclidean
action for the hairy black hole is very close to the Euclidean
action for the topological black hole. In fact, it is slightly
smaller than the latter for $T<\frac{1}{2 \pi} \approx 0.159$ and
slightly larger after that value. If it were the only branch
present, one would thus conclude that the hairy black hole
dominates for small temperatures, while for large temperatures the
topological black hole would be preferred. This would be a
situation similar to the one of the MTZ black hole.

 However, the
two additional branches change completely our conclusions. The
upper branch shows that the hairy black hole dominates up to $T
\approx 0.20.$ When the coupling constant $g$ decreases, equation
(\ref{Tm3}) along with the demand that the temperature should be
positive, show that the acceptable values of $r_0$ are two rather
than three, as may be seen in figure \ref{mass_action_q00005},
left panel. The lowest branch of the previous corresponding figure
\ref{mass_action_q3} shrinks for decreasing $g$ and it finally
disappears. An interesting consequence of this is that the
temperature has an upper limit. The graph for the Euclidean
actions (figure \ref{mass_action_q00005}, right panel) is
influenced accordingly. There are just two branches for the hairy
black hole, rather than three in figure \ref{mass_action_q3} and
the figure ends on its right hand side at $T \approx 1.25.$ The
continuous line represents the Euclidean action for the
topological black hole with the same temperature. Similar remarks
hold as in the previous case, e.g., the phase transition moves to
$T \approx 0.80.$ In addition, the largest value of $r_0$
corresponds to the upper branch of the hairy black hole.

To understand the nature of this phase transition it is
instructive to draw a kind of phase diagram, so that we can spot
which is the dominant solution for a given pair of $g$ and $T.$ We
depict our result in figure \ref{phd}. The hairy solution
dominates below the curve which shows the critical temperature as a function of the coupling constant $g$. The most striking feature of
the graph is that the critical temperature diverges as $g\to 0$.
Thus, it does not converge to the MTZ value
$\fr{1}{2 \pi} \approx 0.159$ at $g=0$.
For even the slightest nonzero values of $g$ the critical
temperature gets extremely large values! This appears to put the
conformal point (MTZ black hole) in a special status within the set of these
hairy black holes. In other words, the restoration of conformal
invariance is not a smooth process, and the MTZ black hole
solution cannot be obtained in a continuous way as $g\rightarrow
0$. In fact, it seems that (even infitesimally) away from the conformal point $g=0$
black holes are mostly hairy!

\begin{figure}[!t]
\centering
\includegraphics[scale=0.3,angle=-90]{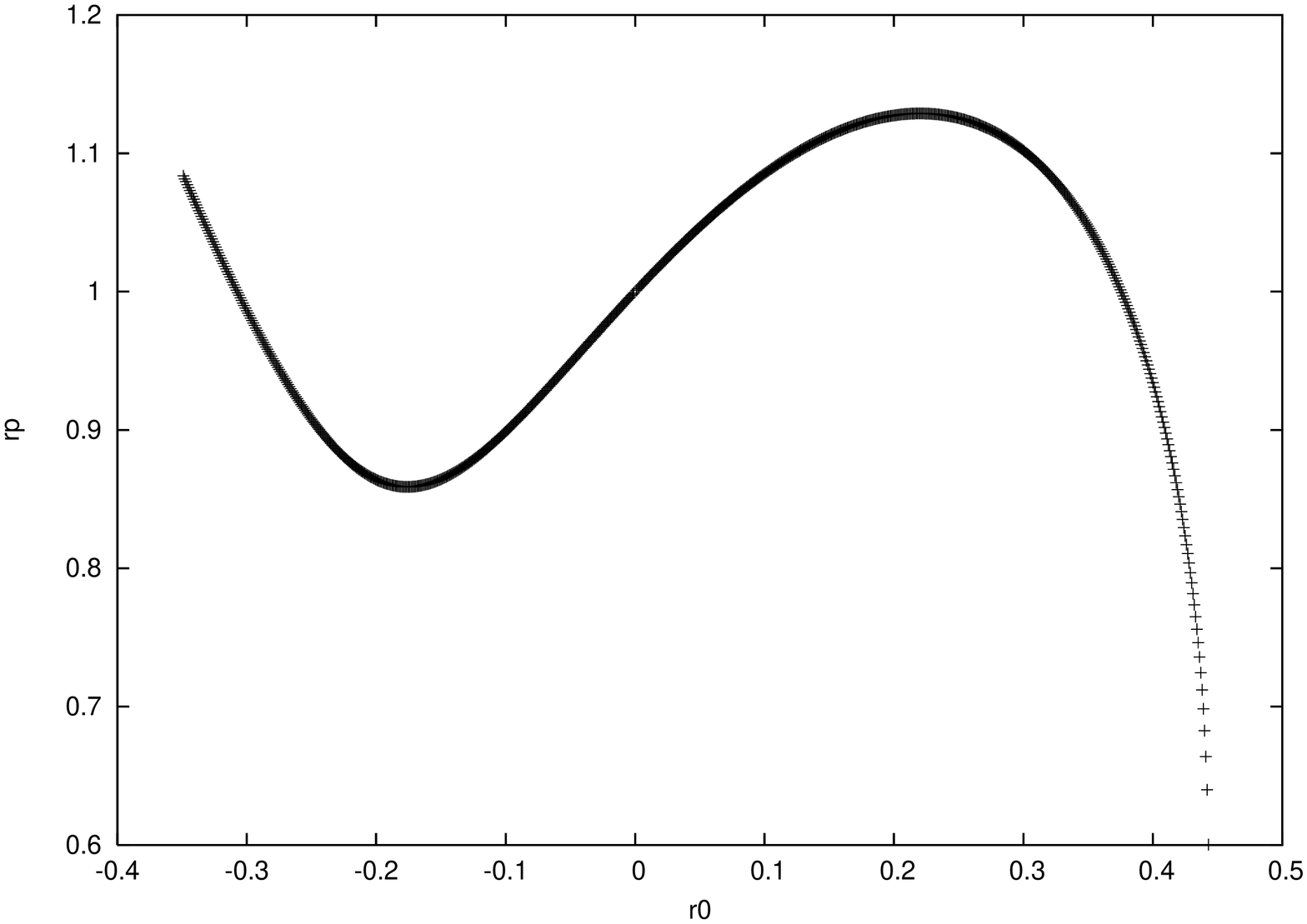}
\includegraphics[scale=0.3,angle=-90]{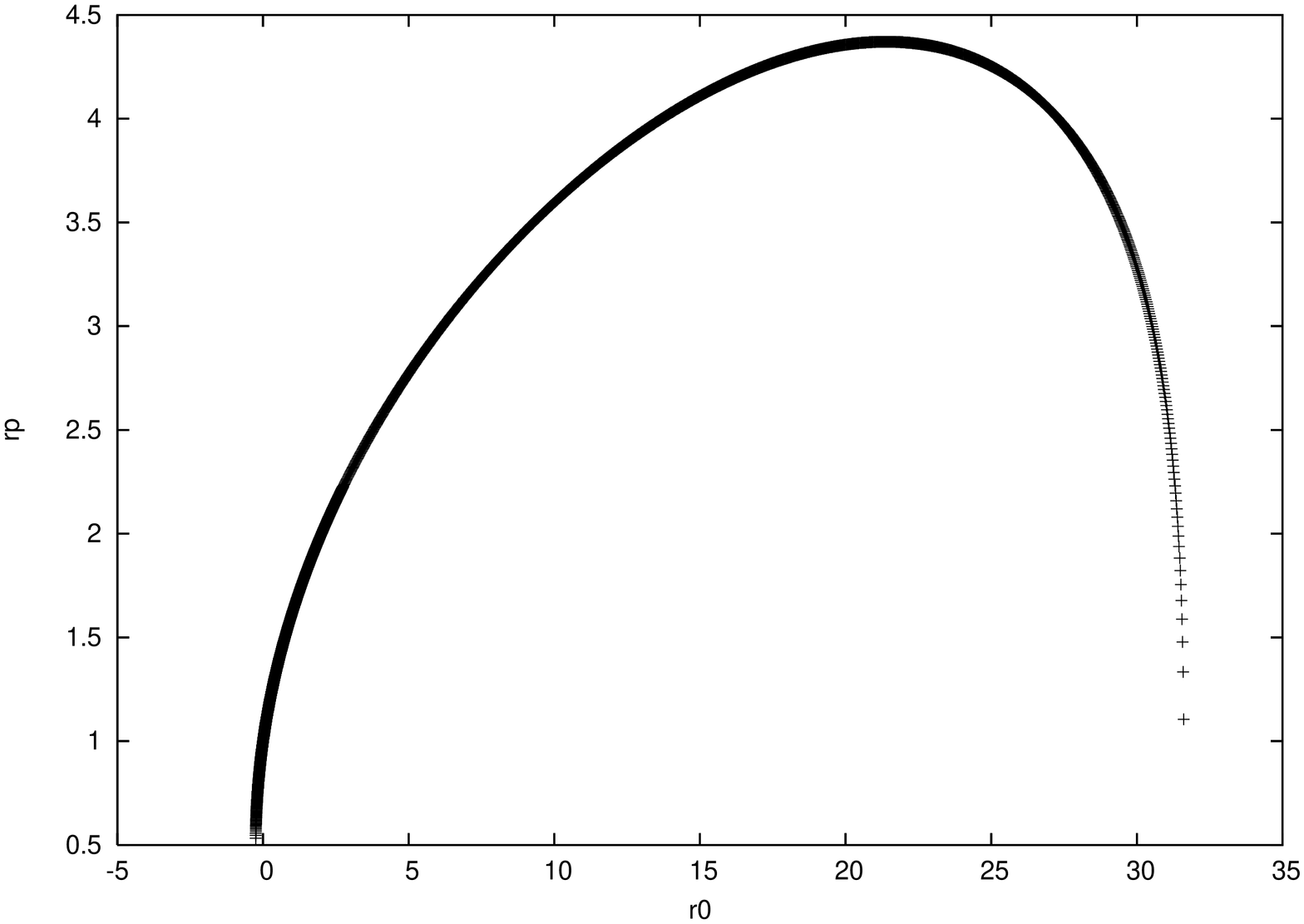}
\caption{Horizon versus parameter $r_0$ for $g=3$ (left
panel) and $g=0.0005$ (right panel).} \label{horizons}
\end{figure}

\begin{figure}[!t]
\centering
\includegraphics[scale=0.3,angle=-90]{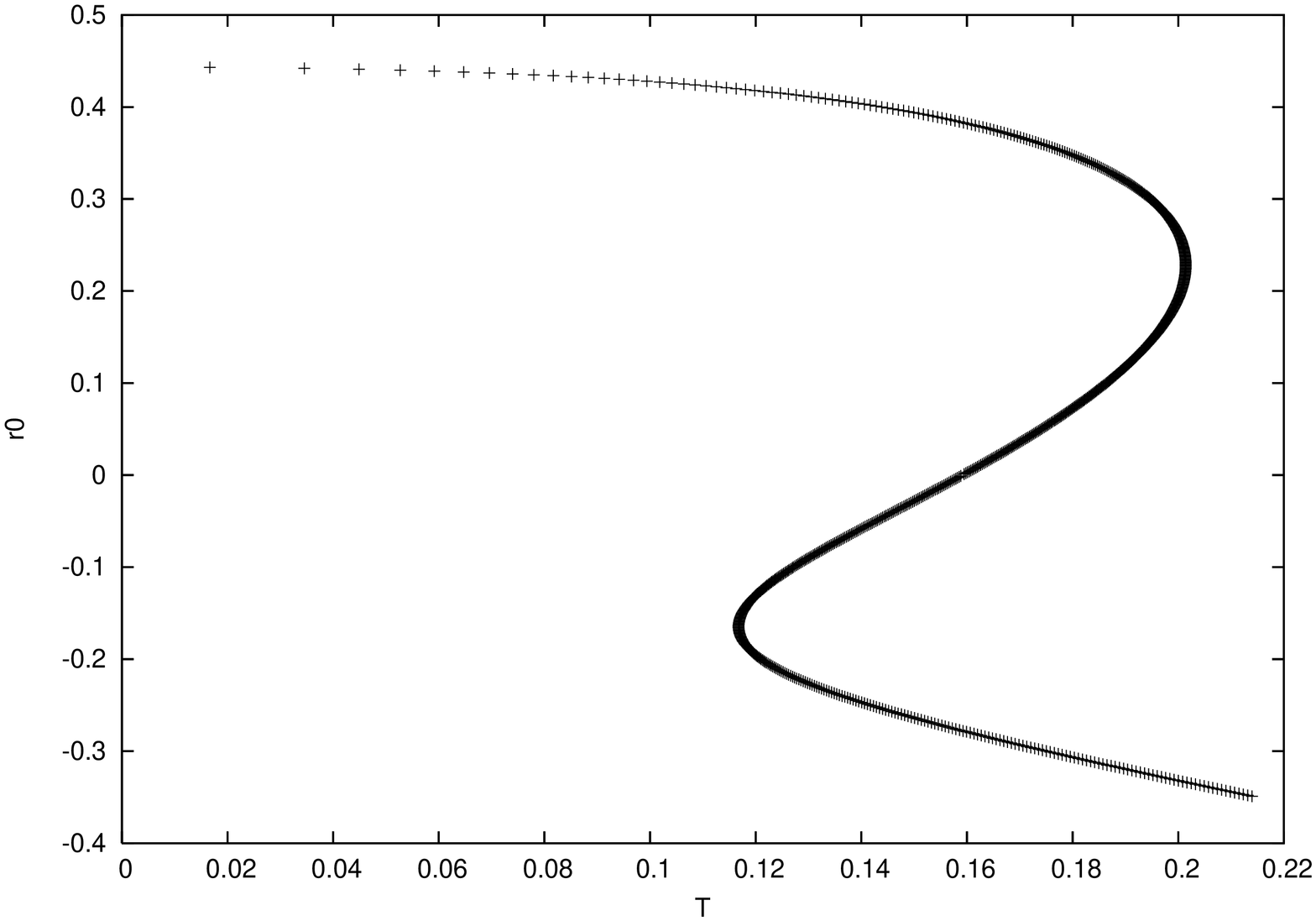}
\includegraphics[scale=0.3,angle=-90]{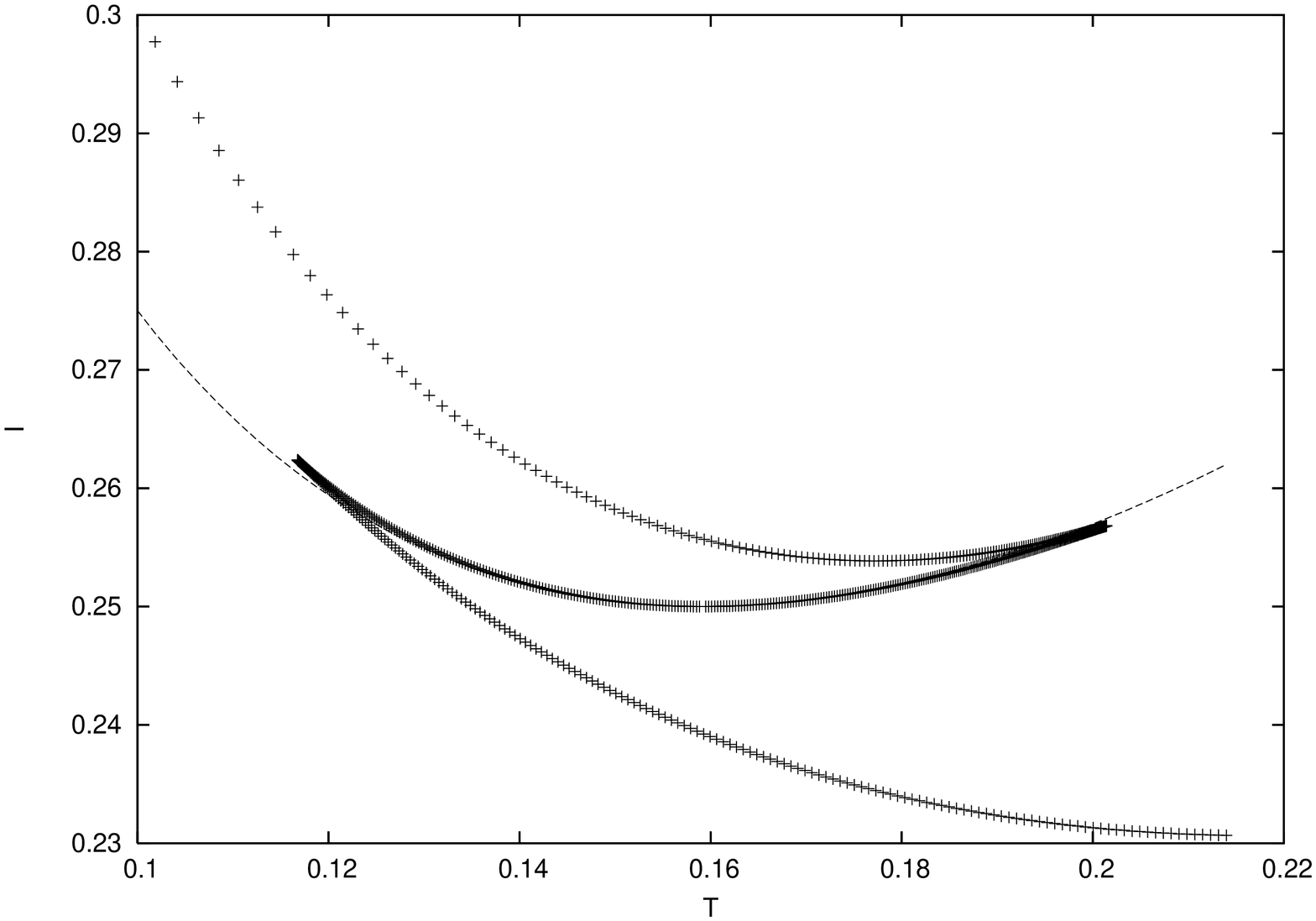}
\caption{Parameter $r_0$ (left panel) and Euclidean actions versus
temperature for $g=3$ (right panel).} \label{mass_action_q3}
\end{figure}

\begin{figure}[!b]
\centering
\includegraphics[scale=0.3,angle=-90]{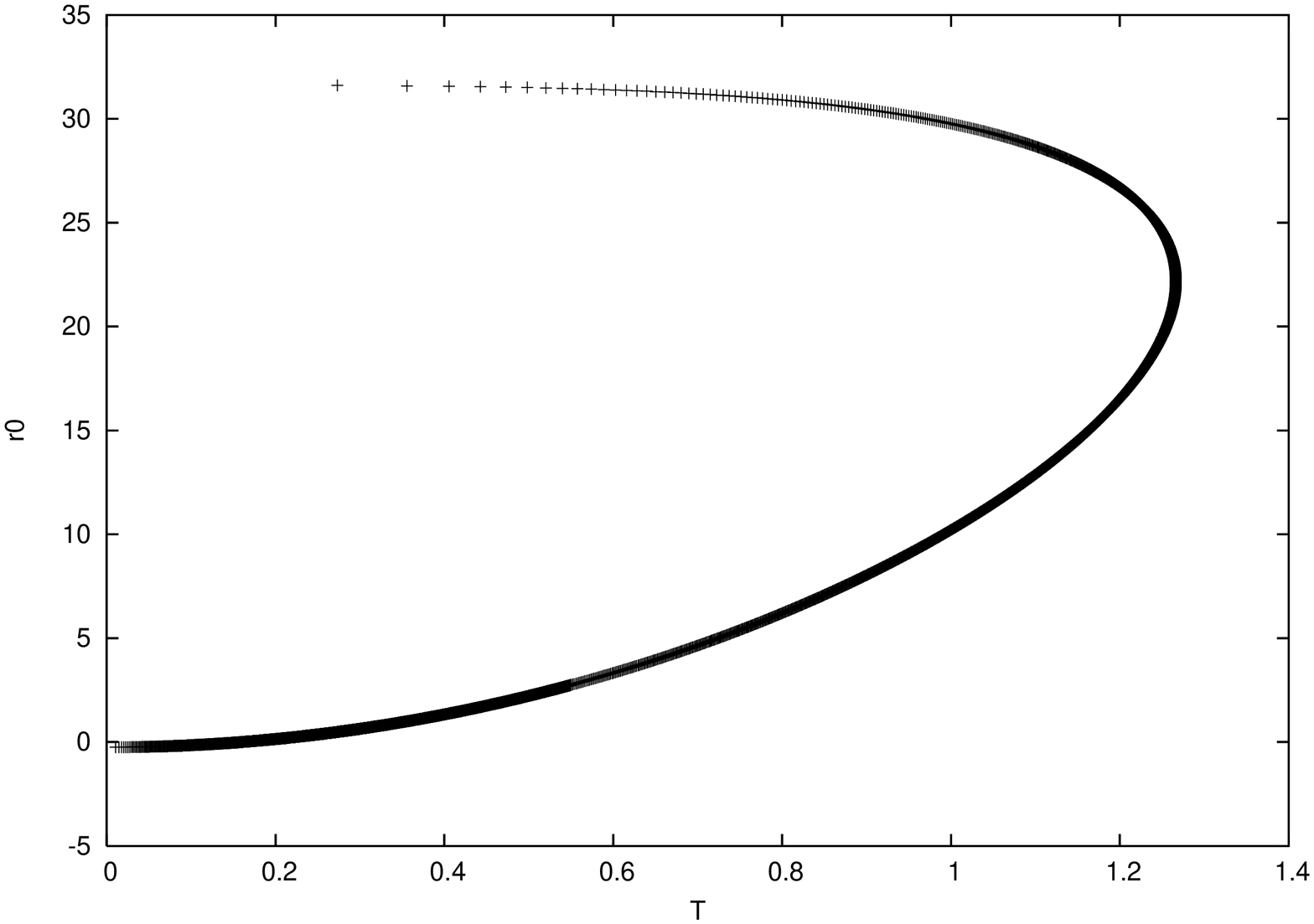}
\includegraphics[scale=0.3,angle=-90]{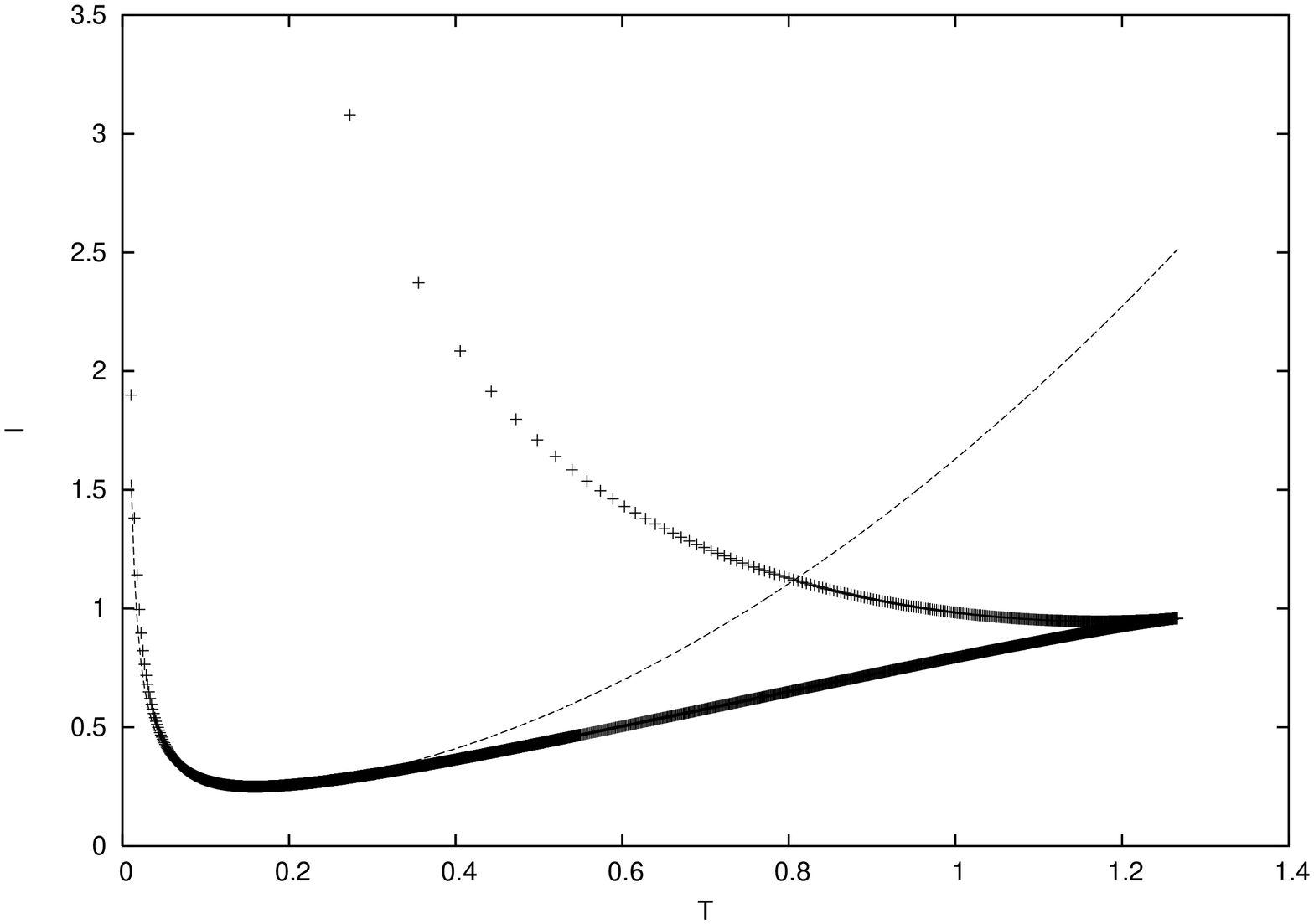}
\caption{Parameter $r_0$ (left panel) and Euclidean actions versus
temperature for $g=0.0005$ (right panel).}
\label{mass_action_q00005}
\end{figure}

\begin{figure}[!t]
\centering
\includegraphics[scale=0.3,angle=-90]{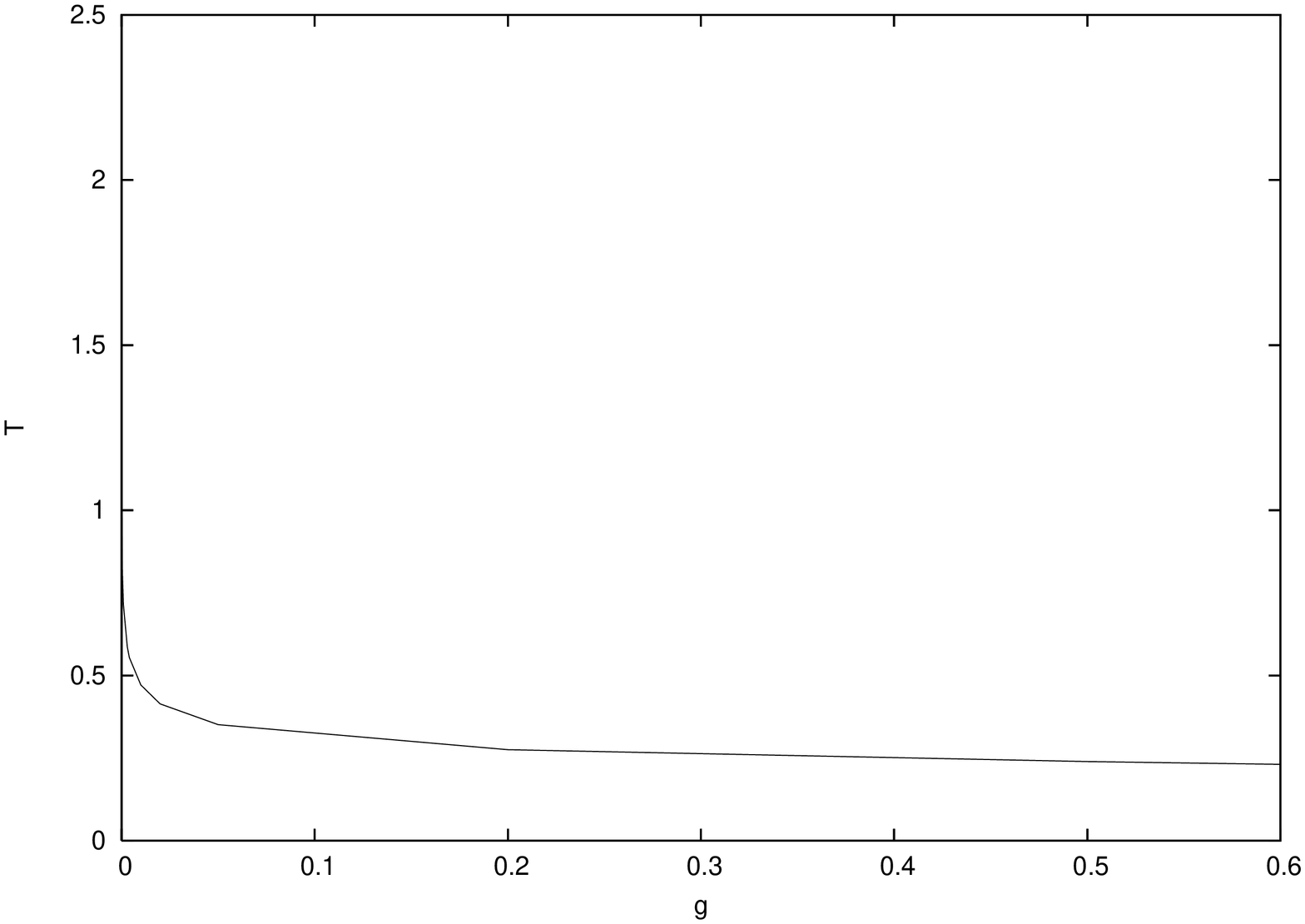}
\caption{Phase diagram. For points under the curve the hairy
solution will be preferred.} \label{phd}
\end{figure}

\section{Stability Analysis}
\label{secst}

To perform the stability analysis of the hairy black hole it is
more convenient to work in the Einstein frame. Henceforth, we
shall work in units in which the radius of the boundary is $l=1$.

We begin with the hairy black hole line element,
\begin{equation}\label{mtzhat} ds_0^2=\frac{\hat{r}(\hat{r}+2 r_0)}{(\hat{r}+r_0)^2} \left[-F(\hat{r})
dt^2+\frac{d\hat{r}^2}{F(\hat{r})}+\hat{r}^2 d \sigma^2\right],
\ee which can be written in the form
\begin{equation}
ds^{2}=-\frac{f_0}{h_0^{2}}dt^{2}+\frac{dr^{2}}{f_0}+r^{2}d\sigma^{2}
\end{equation}
using the definitions \bea f_0(r) = F(\hat{r}) \left( 1 +
\frac{r_0^2}{(\hat{r} + 2r_0)(\hat{r}+r_0)} \right)^2~, \nonumber\\
h_0(r) = \left( 1 +\frac{r_0^2}{(\hat{r} + 2r_0)(\hat{r}+r_0)}
\right) \frac{\hat{r}+r_0}{\sqrt{\hat{r}(\hat{r}+2 r_0)}}~, \eea
under the change of coordinates \bea
r=\fr{\hat{r}^{3/2}(\hat{r}+2r_0)^{1/2}}{\hat{r}+r_0}~. \eea The
scalar field solution reads \be \phi_0 (r) = \sqrt{\frac{3}{4\pi
G}} \tanh^{-1} \frac{r_0}{\hat{r}+r_0}~, \ee obeying the boundary
conditions (\ref{eqBCphi}) with
\be\label{eqxc} \alpha =\alpha_0 = \sqrt{\frac{3}{4\pi G}} |r_0|~.
\ee

We are interested in figuring out when the black hole is
unstable (losing its hair to turn into a TBH) and discuss the results in the context of thermodynamic
considerations.
To this end, we apply the perturbation
\begin{equation}
f(r,t)=f_{0}(r)+f_{1}(r)e^{\omega t} ,\,\,\,\,\,
h(r,t)=h_{0}(r)+h_{1}(r)e^{\omega t},\,\,\,\,\,
\phi(r,t)=\phi_{0}(r)+\frac{\phi_{1}(r)}{r} e^{\omega
t}~.\,\,\label{perts}
\end{equation}
which respects the boundary conditions (\ref{eqBCphi}) with $\omega >0$ for an instability to develop.

The field equations read: \be -1-f-r f^\prime+r f
\fr{h^\prime}{h}+8 \pi G r^2 V(\phi)=0~, \ee \be \dot{f}+r f
\dot{\phi}\phi^\prime=0~,\ee \be 2 h^\prime+r h
\left[\fr{h^2}{f^2}\dot{\phi}^2+\phi^{\prime 2}\right] =0~,\ee \be
\left(\fr{\dot{h}}{f} \dot{\phi}\right)-\fr{1}{r^2}\left(r^2
\fr{f}{h} \phi^\prime \right)^\prime + \fr{1}{h}
V^\prime(\phi)=0~.\ee

The field equations give a Schr\"odinger-like wave equation for
the scalar perturbation,
\begin{equation}
-\frac{d^{2}\phi_{1}}{dr^{2}_{*}}+\mathcal{V}\phi_{1}=-\omega^{2}\phi_{1}~,
\label{scal1}
\end{equation}
where we defined the tortoise coordinate \be
\frac{dr_{*}}{dr}=\frac{h_{0}}{f_{0}}~, \ee and the effective
potential is given by \be \mathcal{V} = \frac{f_0}{h_0^2} \left[ -
\frac{1}{2} (1+r^2 {\phi_0'}^2) {\phi_0'}^2 f_0 + (1-r^2
{\phi_0'}^2 ) \frac{f_0'}{r} + 2r\phi_0' V'(\phi_0) + V''(\phi_0)
\right]. \ee The explicit form of the Schr\"odinger-like equation
reads: \be -F(\hat{r}) \fr{d}{d \hat{r}}\left[F(\hat{r})
\fr{d\phi_1}{d \hat{r}}\right]+{\cal V} \phi_1 = -\omega^2 \phi_1,
\ee where the functional form of the function $F$ has been given
in equation (\ref{Psiconf}) and \footnote{We have set $8\pi G=1$
.}

\be {\cal V} = \fr{r_0^2 F(\hat{r})}{\hat{r}^2 \left(1+\fr{2
r_0}{\hat{r}}\right)^2 \left(1+\fr{3 r_0}{\hat{r}}+\fr{3
r_0^2}{\hat{r}^2}\right)^2}\left\{5 + \fr{2+(11 g+54)r_0^2}{r_0
\hat{r}} + \fr{29+(47 g+189) r_0^2}{\hat{r}^2} \right. \ee \be
\left. +\fr{r_0 (150+(-3 g+270) r_0^2)}{\hat{r}^3}+ \fr{r_0^2
(396+(-351 g+135) r_0^2)}{\hat{r}^4}+ \fr{r_0^3 (612 -873 g
r_0^2)}{\hat{r}^5} \right. \ee \be \left. +\fr{r_0^4 (582-1047 g
r_0^2)}{\hat{r}^6} + \fr{324 r_0^5 (1 - 2 g r_0^2)}{\hat{r}^7} +
\fr{81 r_0^6 (1 - 2 g r_0^2)}{\hat{r}^8}   \right. \ee \be \left.
+\fr{g}{2} \left(5+\fr{54 r_0}{\hat{r}} + \fr{189
r_0^2}{\hat{r}^2} +\fr{270 r_0^3}{\hat{r}^3} +\fr{135
r_0^4}{\hat{r}^4} \right) \ln\left(1+\fr{2 r_0}{\hat{r}}\right)
\right\}~.\ee

Near the horizon the Schr\"odinger-like equation simplifies to \be
-[f^\prime(\hat{r}_+)]^2 \epsilon \fr{d}{d \epsilon}
\left[\epsilon \fr{d \phi_1}{d \epsilon} \right] = -\omega^2
\phi_1, \ \ \epsilon = \hat{r}-\hat{r}_+ \ee and its acceptable
solution reads \be \phi_1 \sim \epsilon^{\kappa \omega}, \ \ \
\kappa=\fr{1}{f^\prime(\hat{r}_+)}, \ \ \omega>0~. \ee Regularity
of the scalar field at the horizon ($r\to r_+$) requires the
boundary conditions \be \phi_1 = 0 \ \ , \ \ \ \ (r-r_+) \phi_1' =
\kappa\omega \phi_1\ \ , \ \ \ \ \kappa
> 0 ~. \ee For a given $\omega >0$, they uniquely determine the
wavefunction.

At the boundary $(\hat{r} \to \infty),$ the wave equation is
approximated by \be - \frac{d^2\phi_1}{dr_*^2} + 5r_0^2 \phi_1 =
-\omega^2 \phi_1~, \ee with solutions \be \phi_1 = e^{\pm E r_*} \
\ , \ \ \ E = \sqrt{\omega^2 + 5 r_0^2}~, \ee where $ r_* = \int
\fr{d \hat{r}}{f(\hat{r})}= - \frac{1}{r} + \dots $~. Therefore,
for large $r$, \be \phi_1 = A + \frac{B}{r} + \dots
\label{fitAB}\ee To match the boundary conditions (\ref{eqBCphi}), we need \be\label{eqx33} \frac{B}{A} = 2 c \alpha_0
=- 2 r_0~. \ee
Since
the wavefunction has already been determined by the boundary
conditions at the horizon and therefore also the ratio $B/A$, this
is a constraint on $\omega.$ If (\ref{eqx33}) has a solution, then
the black hole is unstable. If it does not, then there is no
instability of this type (however, one should be careful with
non-perturbative instabilities).

In figure \ref{stable_q0_q00005} (left panel) we show the ratio
$B/A$ for the standard MTZ black hole (corresponding to $g=0$)
versus $\omega$ at a typical value of the mass parameter, namely
$r_0=-0.10.$ It is obvious that the value of the ratio lies well
below the values $2 r_0=+0.20.$ It is clearly impossible to have a
solution to this equation, so the solution is stable. In fact this
value of the mass parameter lies in the interesting range for this
black hole, since thermodynamics dictates that for negative values
of $r_0$ MTZ black holes are favored against  topological  black
holes. We find that the MTZ black holes turn out to be stable.

Next we examine the case $g=0.0005,$ for which we have presented
data before in figure \ref{mass_action_q00005}. As we have
explained there, the most interesting branch of the graphs is the
upper branch, on the right panel, which dominates the small $T$
part of the graph and corresponds to large values of $r_0$
(typically around $30$ on the left panel of the same figure). Thus
we set $g=0.0005, \ \ r_0=+30$ and plot $B/A$ versus $\omega$ in
the right panel of figure \ref{stable_q0_q00005}. It is clear that
the curve lies systematically below the quantity $-2 r_0 = -60$
and no solution is possible, so the hairy black hole with these
parameters is stable.

Finally we come to the case with $g=+3,$ which has three branches.
In the left panel of figure \ref{stable_q3} we show the results
for $r_0=+0.40,$ which corresponds to the upper branch of figure
\ref{mass_action_q3}; the curve again lies below the quantity $-2
r_0 = -0.80$ and no solution is possible, so this hairy black hole
is stable. In the right panel of figure \ref{stable_q3} we show
the results for $r_0=-0.30,$ which corresponds to the lowest
branch of figure \ref{mass_action_q3}, which disappears for
decreasing $g.$ In this case we find something qualitatively
different: the curve cuts the line $-2 r_0 = +0.60$ around $\omega
\approx 0.40$ and a solution is possible, signaling instability.
Thus, for $g=3$ the hairy black hole may be stable or unstable,
depending on the value of $r_0.$

\begin{figure}[!t]
\centering
\includegraphics[scale=0.3,angle=-90]{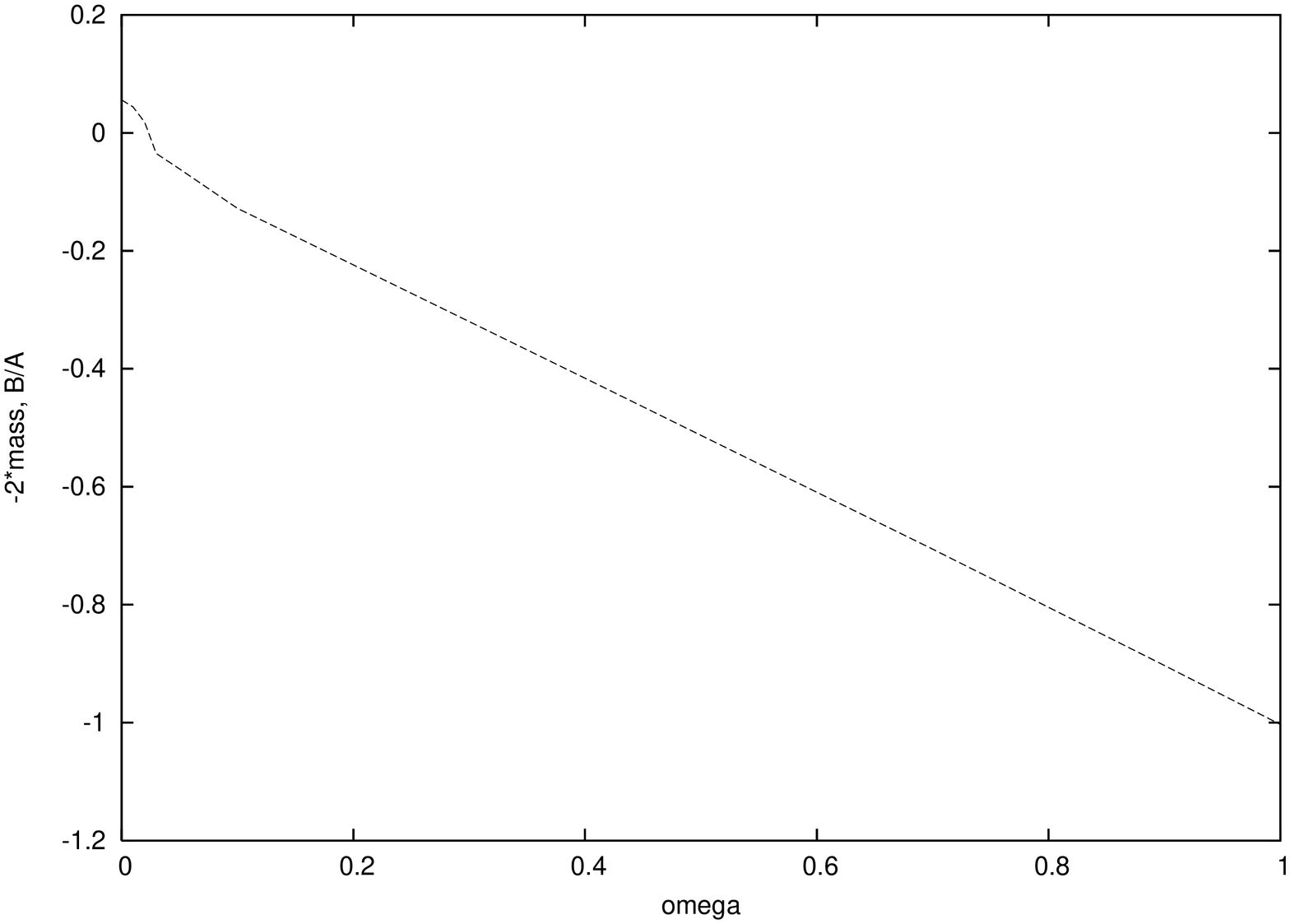}
\includegraphics[scale=0.3,angle=-90]{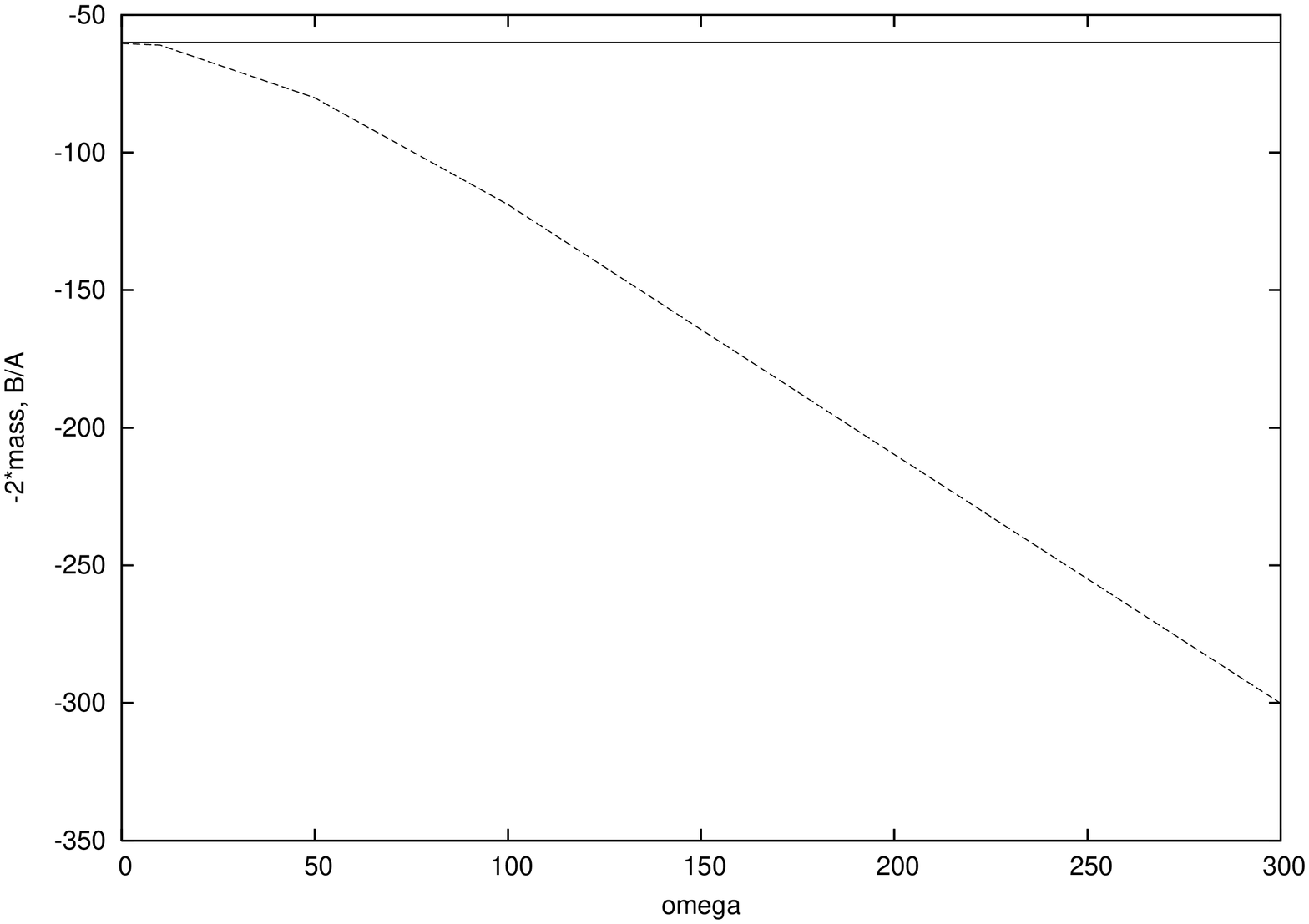}
\caption{Stability of the standard MTZ black hole (left panel) for
$r_0=-0.10;$ similarly for the hairy black hole at $g=0.0005, \
r_0=+30$ (right panel).} \label{stable_q0_q00005}
\end{figure}

\begin{figure}[!t]
\centering
\includegraphics[scale=0.3,angle=-90]{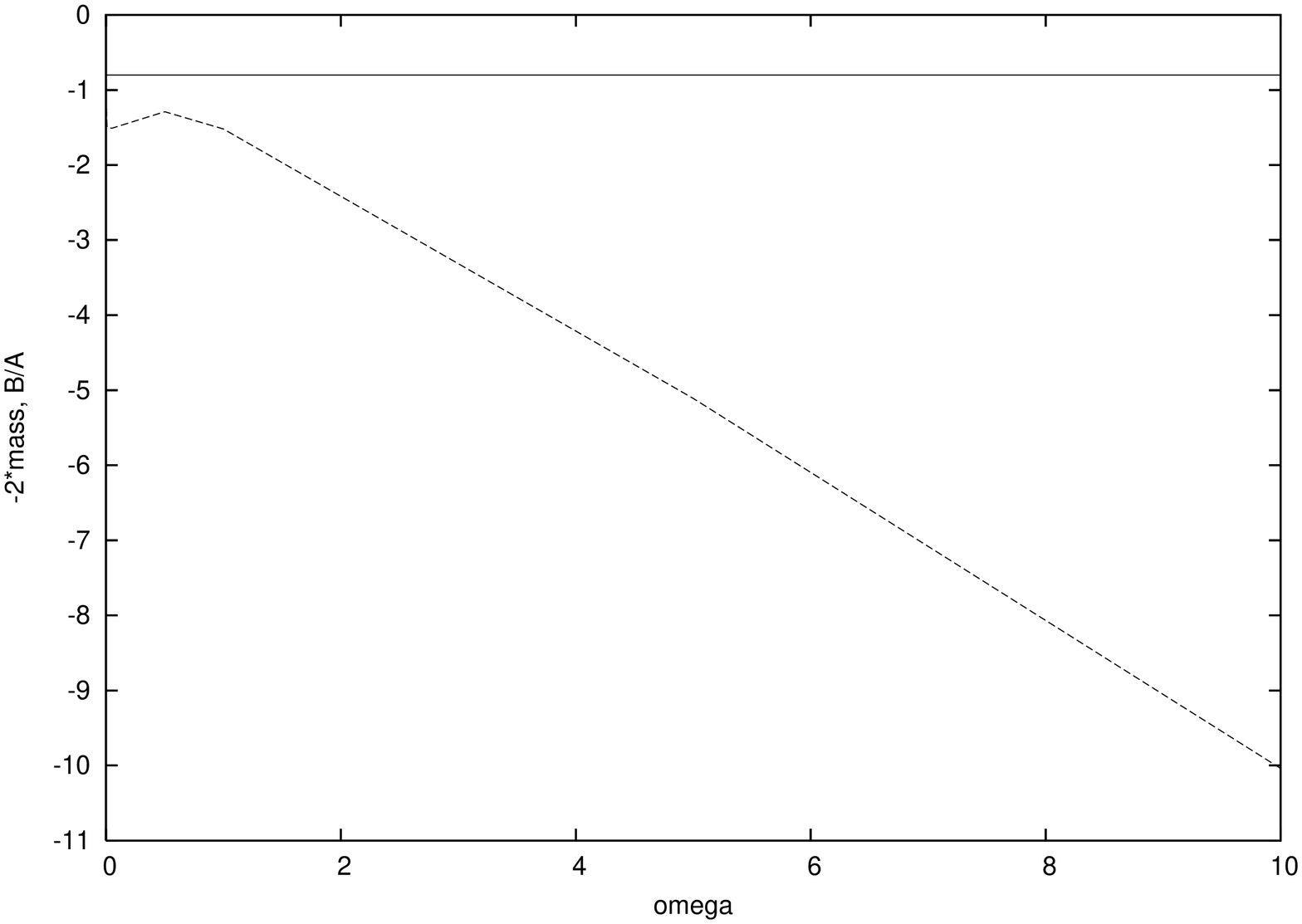}
\includegraphics[scale=0.3,angle=-90]{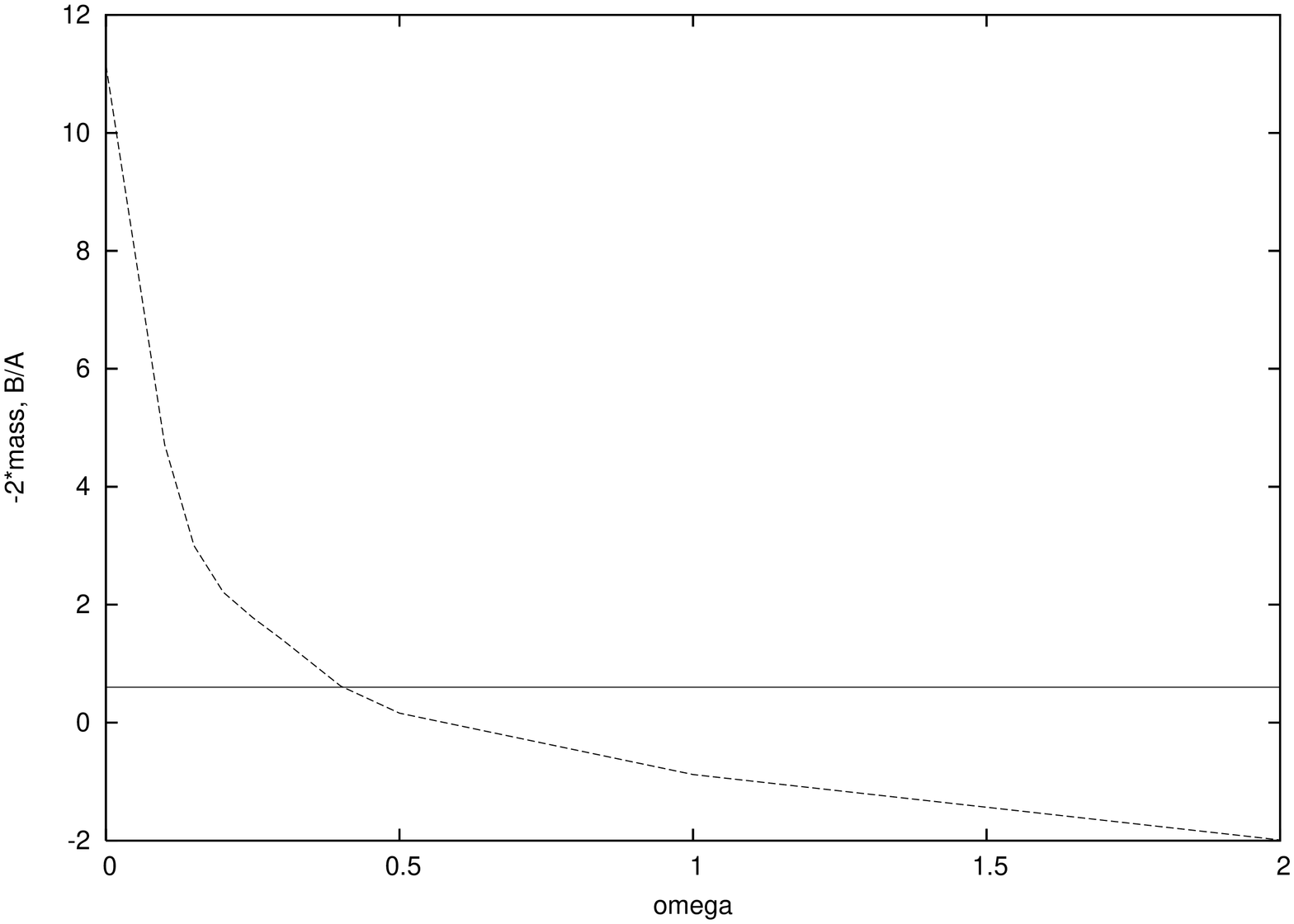}
\caption{Stability of the hairy black hole  for $g=3$ and
$r_0=+0.40$ (left panel); similarly for the hairy black hole at
$r_0=-0.30$ (right panel).} \label{stable_q3}
\end{figure}

\section{Conclusions}
\label{sec6}

We presented a new class of hairy black hole solutions in
asymptotically AdS space. The scalar field is minimally coupled to
gravity with a non-trivial self-interaction potential. A coupling
constant $g$ in the potential parametrizes our solutions. If $g=0$
the conformal invariant MTZ black hole solution, conformally
coupled to a scalar field, is obtained. If $g\neq 0$ a whole new
class of hairy black hole solutions is generated. The scalar field
is conformally coupled but the solutions are not conformally
invariant. These solutions are perturbative stable near the
conformal point for negative mass and they may develop
instabilities for positive mass.

We studied the thermodynamical properties of the solutions.
Calculating the free energy we showed that for a general $g$,
apart the phase transition of the MTZ black hole at the critical
temperature $T=1/2\pi l$, there is another critical temperature,
higher than the MTZ critical temperature, which depends on $g$ and
where a first order phase transition occurs of the vacuum black
hole towards a hairy one. The existence of a second critical
temperature is a manifestation of the breaking of conformal
invariance. As $g\rightarrow 0$ the second critical temperature
diverges, indicating that there is no smooth limit to the MTZ
solution.

The solutions presented and discussed in this work have hyperbolic
horizons. There are also hairy black hole solutions with flat or
spherical horizons of similar form. However, these solutions are
pathological. In the solutions with flat horizons, the scalar
field diverges at the horizon, in accordance to the ``no-hair"
theorems. In the case of spherical horizons, calculating the free
energy we find that always the vacuum solution is preferable over
the hairy configuration. Moreover, studying the asymptotic
behaviour of the solutions, we found that they are unstable for
any value of the mass.

\section*{Acknowledgments}

G.~S.~was
supported in part by the US Department of Energy under grant
DE-FG05-91ER40627.

\end{document}